\documentclass[prl,showpacs,amsfonts,twocolumn]{revtex4-1}
\usepackage{amsmath,amssymb,graphicx,hyperref, color,amsbsy}
\hypersetup{%
  pdfauthor={Jonathan Keeling},
  pdftitle={Superfluid density of an open dissipative condensate}}
\renewcommand{\vec}[1]{\boldsymbol{#1}}
\newcommand{\Tr}{\text{Tr}}
\newcommand{\ZZ}{\mathcal{Z}}
\newcommand{\nodagger}{{\vphantom{\dagger}}}
\begin{document}

\title{Superfluid density of an open dissipative condensate}
\author{Jonathan Keeling} 
\affiliation{SUPA, School of Physics and Astronomy, University of St Andrews, 
  KY16 9SS, UK}

\date{\today}
\begin{abstract}
  I calculate the superfluid density of a non-equilibrium steady state
  condensate of particles with finite lifetime.  Despite the absence
  of a simple Landau critical velocity, a superfluid response
  survives, but dissipation reduces the superfluid fraction. I also
  suggest an idea for how the superfluid density of an example of
  such a system, i.e.\ microcavity polaritons, might be measured.
\end{abstract}

\maketitle

The observation of superfluidity is one of the most compelling
signatures of quantum coherence in systems such as Helium and cold
atomic gases~\cite{leggett06}.  Recently, there has been much activity
exploring condensation of mixed matter-light excitations, i.e.\
semiconductor microcavity polaritons (see~\cite{deng10} for a review)
as well as recent experiments on photons in dye filled
cavities~\cite{Klaers10}.  Microcavity polaritons consist of
superpositions of excitons confined in quantum wells, and photons
confined in semiconductor microcavities.  Because of the photonic
component of these particles, they have a finite lifetime, and so any
condensate will be a non-equilibrium steady state, where loss is
balanced by injection of new particles.  As well as these matter-light
systems, recent experiments on continuous loading of cold atoms
into traps~\cite{Falkenau2011} suggest that similar questions of
superfluid properties in non-equilibrium steady states may soon be
accessible in cold atoms.

Finite particle lifetime prompts questions about the meaning of
superfluidity when the particles involved in the condensate are
continually being replaced.  To see why such questions arise, one may
observe that finite particle lifetime changes the excitation spectrum
from the linearly dispersing Bogoliubov sound mode to a diffusive
mode~\cite{szymanska06,wouters06}, $\xi(k) = - i \sigma + \sqrt{c^2
  k^2 - \sigma^2}$.  The form of the spectrum of long wavelength modes
is commonly invoked in explaining the existence of superfluidity,
i.e.\ if the low energy excitations are linearly dispersing Bogoliubov
sound modes, then there exists a critical velocity, such that for a
fluid flowing below this velocity, it is not energetically favorable
to create quasiparticle excitations, and thus the flowing superfluid
remains stable~\cite{leggett06}.  If one naively defines a critical
velocity by the dispersion of the real part of $\omega(k)$, one finds
that for the non-equilibrium spectrum, such a critical velocity
vanishes.  As has been pointed out
elsewhere~\cite{wouters10:superfluid,wouters10:vortices}, a more
careful argument regarding the response to a static defect indicates
that there may still be a particular velocity at which there is a
sharp onset of drag.  Nonetheless, the above illustrates why
non-equilibrium condensates require one to re-examine questions of
superfluidity.

Experimentally, aspects of superfluid behavior have been studied in
microcavity polariton systems, and features such as quantized
vortices~\cite{Lagoudakis2008}, suppression of scattering off
disorder~\cite{Amo2009,*Amo2009a} and metastability of induced
vortices~\cite{Krizhanovskii2010}  have been observed.
While these various experiments show that aspects of superfluid
behavior can be seen in such non-equilibrium condensates, they leave
open an important question, namely what is the superfluid fraction.
Even in superfluid Helium, at non-zero temperatures there are drag
forces due to the normal fluid component.  However, the normal and
superfluid components can be clearly distinguished by their response
to a slow rotation~\cite{leggett06}: At low angular velocities, the
superfluid cannot rotate, so only the normal component rotates, thus
reducing the effective moment of inertia.  Determining the normal and
superfluid densities thus gives a fuller description of the superfluid
properties of a system than does a binary distinction between superfluid
and non-superfluid systems.  As such, the aim of this letter is to
discuss the normal and superfluid densities of an open dissipative
condensate, and to suggest how this might be explored in the
microcavity polariton systems.

The same superfluid density as introduced above can be found from the
current-current response function $\chi_{ij}(q)$.  This response
function gives the particle current $J_i(q) = \sum_k
\psi^\dagger_{k+q} \gamma_i(2\vec{k}+\vec{q}) \psi_k$ (where the
current vertex $\gamma_i(\vec{k})= k_i/2m$ and $\hbar=1$ throughout) due to a perturbation $ H
\to H - \sum_q f_i(q) J_i(q)$, i.e. $J_i(q) = \chi_{ij}(q) f_j(q)$.
In an isotropic system, one may then define $\rho_s, \rho_n$
as~\cite{leggett06}
\begin{equation}
  \label{eq:2}
  m \chi_{ij}(q \to 0,\omega=0) = \rho_s \frac{q_i q_j}{q^2} + \rho_n \delta_{ij}.
\end{equation}
The superfluid component picks out and responds only to the
irrotational (non-transverse) part of the applied force.  The
advantage of this approach is that it allows an explicit calculation
for the open dissipative system, using the Schwinger-Keldysh approach
for non-equilibrium systems (see
e.g.~\cite{altland_simons,*kamenev05}).

In the following, I will consider the simplest model of a weakly
interacting dilute Bose gas:
\begin{equation}
  \label{eq:3}
  H = \sum_k \epsilon_k \psi^\dagger_k \psi^\nodagger_k + \sum_{k,k^\prime,q}
  \frac{U}{2} \psi^\dagger_{k+q} \psi^\dagger_{k^\prime-q} \psi^\nodagger_{k^\prime} \psi^\nodagger_k
\end{equation}
where $\epsilon_k = k^2/2m$. In addition, one must include pump and
decay processes such that the bare inverse retarded Green's function
$[D^{R}_{(0)}]^{-1} = \omega - \epsilon_k + i \kappa - i p(\omega)$
where the pump has the form $p(\omega) = \gamma - \eta \omega$ and
$\kappa$ describes decay\footnote{To avoid ultraviolet divergence, a
  regularization $[\kappa - i p(\omega)] \to [\kappa - i p(\omega)] i
  \Gamma/(\omega+i\Gamma)$ is also required.  $\Gamma$ is assumed
  large compared to all other energy scales.}.  This form of pumping
is motivated by recent works by~\citet{wouters10:superfluid} (as well
as related models~\cite{wouters10:relax}), and simplifies the
calculation of $\chi_{ij}$ compared to models with density-dependent
pump processes.  When condensed, such a model has the diffusive
spectrum $\zeta(k) = - i \sigma + \sqrt{c^2 k^2 - \sigma^2}$ with
$\sigma = \eta \mu/(1+\eta^2), c=\sqrt{\mu/[m(1+\eta^2)]}$. Thus,
finding a non-zero superfluid density in such a model addresses how
the diffusive spectrum affects superfluidity.

\begin{figure}
  \centering
  \includegraphics[width=3.2in]{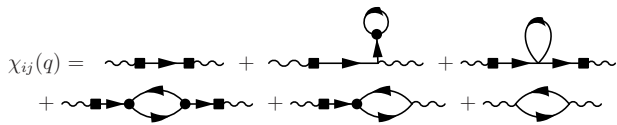}
  \caption{Types of Feynman diagram required for the response function
    to one-loop order.  Straight lines indicate non-condensate
    excitations. Filled symbols involve the condensate, arising from
    either interactions (circles) or coupling to currents (squares).
    Wavy lines indicate source fields coupling to the current
    vertices.}
  \label{fig:feynman}
\end{figure}

To correctly find the superfluid response function~\cite{griffin94}
requires vertex corrections in the response function. In equilibrium,
this can be avoided by using sum rules that result from conservation
of density, however with finite particle lifetimes, this is not
necessarily a-priori justified.  I will postpone until later the
discussion of how the vertex corrections are to be determined and next
summarize the physical reason that a superfluid density can survive.

Fig.~\ref{fig:feynman} illustrates the classes of Feynman diagram
(including vertex corrections) that result at one loop order.  The
first five diagrams contribute to the superfluid density, while the
last gives the normal density.  This can be seen by noting that the
first five diagrams all have the current vertex scatter a particle out
of the condensate, and thus involve a factor $\gamma_i(\vec{q}) \propto
q_i$, hence they all contribute to $\chi_{ij} \propto q_i q_j$.  In
order that the superfluid density does not vanish, it is crucial that
the fluctuation propagator $D^R(\vec{q},\omega=0)$ that also appears in these
five diagrams behave as $1/q^2$ at $q \to 0$ so that overall
$\chi_{ij} \propto q_i q_j/q^2$ remains finite.  The existence of
superfluid density therefore depends on how the denominator of the
Green's function behaves.

In thermal equilibrium, the Green's function behaves as
$D^R(\vec{q},\omega) \propto [(\omega+i0)^2 - c^2 q^2]^{-1}$ and so
the correct scaling of $D^R(\vec{q},\omega=0)$ is dependent on having the a
linear spectrum, hence the relation of the Landau critical velocity
and superfluid density.  However, despite the changed spectrum of the
open dissipative system, one has $D^R(\vec{q},\omega) \propto
[\omega^2 + 2i\sigma\omega - c^2 q^2]^{-1}$ and so the Green's
function at $\omega=0$ still scales as $D^R \propto 1/q^2$, yielding a
non-vanishing superfluid density.  Such behavior of the Green's
function has also been seen to exist in several other models of
non-equilibrium polariton condensates~\cite{szymanska06,wouters06}.
The fact that this structure of the Green's function leads to a
superfluid density, despite the modified spectrum, is the first main
result of this letter.

\begin{figure}
  \centering
  \includegraphics[width=3.2in]{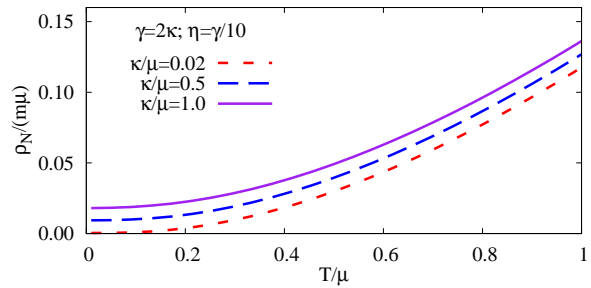}
  \caption{Normal density vs temperature for a variety of pump and
    decay rates and obeying the equilibrium fluctuation
    dissipation theorem.  Relative pump and decay rates are scaled
    so all three curves correspond to the same condensate
    density, while the influence of pump and decay varies.}
  \label{fig:rhon}
\end{figure}

A second result is the effect of finite particle lifetime on the
normal density.  In an equilibrium single component system, the normal
density vanishes at zero temperature~\cite{leggett06}.  The normal
density of the non-equilibrium system can be straightforwardly
calculated since, just as in the thermal equilibrium case, there are
no vertex corrections at one loop order~\cite{griffin94}, so one finds (in
2D):
\begin{equation}
  \label{eq:4}
  \frac{\rho_n}{m} = - \iint \frac{d\epsilon_k}{2\pi} \frac{d\omega}{2\pi}
  \epsilon_k  \frac{i}{4}
  \Tr\left[\sigma_3 D^K_{k} \sigma_3 (D^R_{k} + D^A_{k}) \right]
\end{equation}
where the Green's functions and Pauli matrices $\sigma_i$ are written in
Nambu space, i.e.\ $D^R_k(t,t^\prime) = -i \theta(t-t^\prime)\langle
[\Psi^\nodagger_k(t),\Psi^\dagger_k(t^\prime)] \rangle, \Psi^{\dagger}_k =
(\psi^\dagger_k, \psi^\nodagger_{-k})$.  Even for a thermalised case, using
the equilibrium fluctuation dissipation theorem, $D^K_k(\omega)
= (2n_B(\omega) + 1) [D^R_k(\omega) -D^A_k(\omega)]$, one finds that the
presence of pump and decay terms affect the normal density. As shown
in Fig.~\ref{fig:rhon}, the normal density does not vanish 
at zero temperature.

Having shown that superfluid density need not vanish in a dissipative
condensate, but is reduced by finite lifetime, one may then ask how
the superfluid and normal densities could be measured in such a
system.  As an illustration, the following suggests a method 
that uses  the polariton polarization degree of
freedom~\cite{shelykh10:review} in order to apply ideas that have only
recently been proposed for how one might  measure of
superfluid density in cold atom
systems~\cite{Cooper2010,dalibard,*Cooper2011}.  A number of
alternative methods likely also exist, such as adaptations of
proposals to create gauge fields in coupled photonic
cavities\cite{Hafezi2007,*Koch2010,*Umucalilar2011}.
To adapt the approach in~\cite{dalibard,*Cooper2011}, one first considers the
effective Hamiltonian $H_B$ due to an inhomogeneous real magnetic
field, in the space of polariton polarization states.  If the
splitting of polarization states is always large, one may restrict to
the adiabatic ground state $ |\psi\rangle$.  Because the polarization
composition of this ground state varies in space, there can be a
non-trivial synthetic gauge field in this subspace $q
\vec{A}_{\text{synth}} = i \langle \psi|\nabla \psi \rangle$.  Thus, a
real magnetic field acting on polarization degrees of freedom can
induce an artificial vector potential acting on the (neutral)
polaritons.  This artificial gauge field can mimic a rotating frame,
thereby allowing one to distinguish the superfluid and normal response
to rotation~\cite{leggett06}.

\begin{figure}
  \centering
  \includegraphics[width=1.3in]{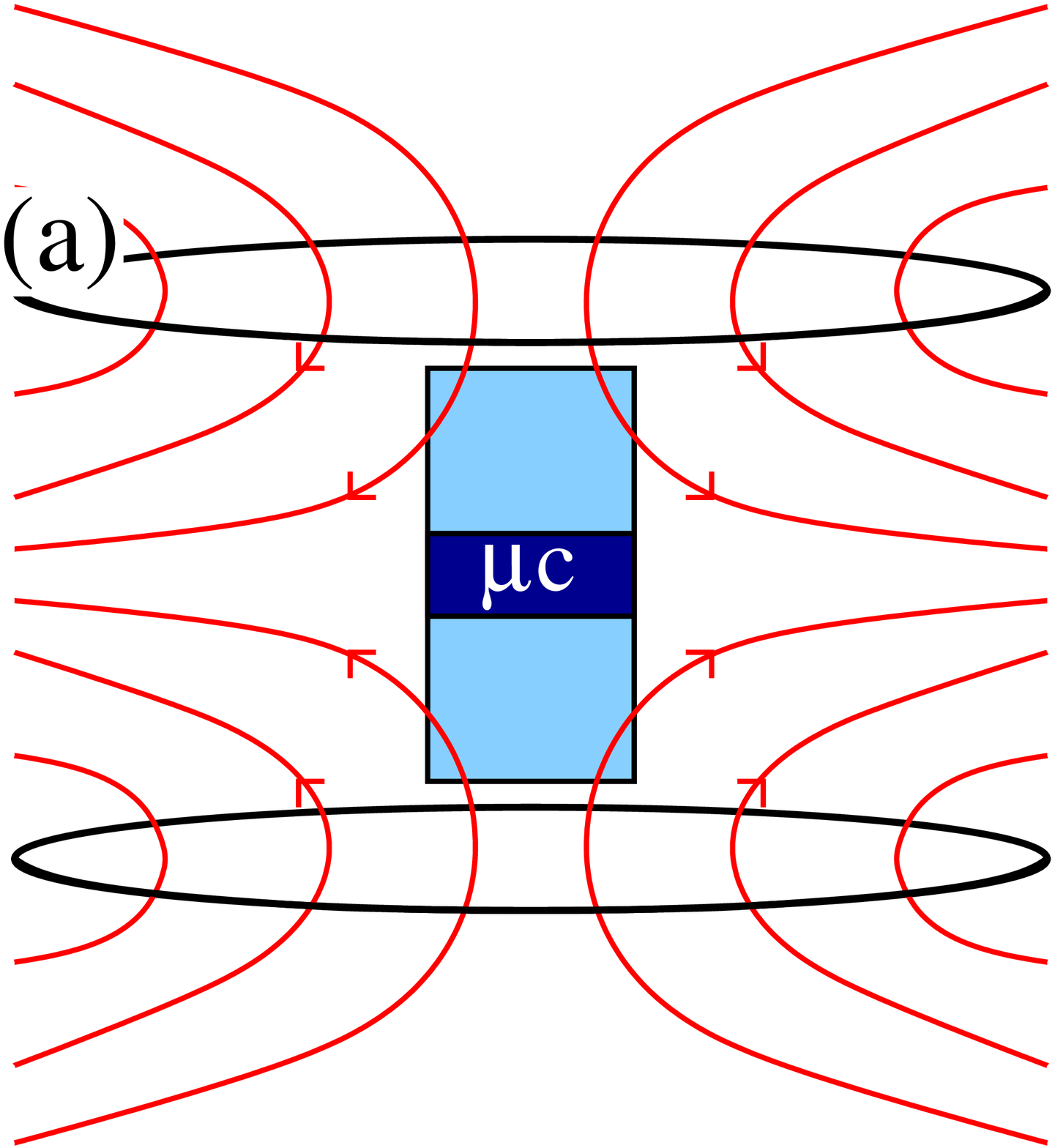}
  \hspace{0.1in}
  \raisebox{-0.22in}{\includegraphics[width=1.55in]{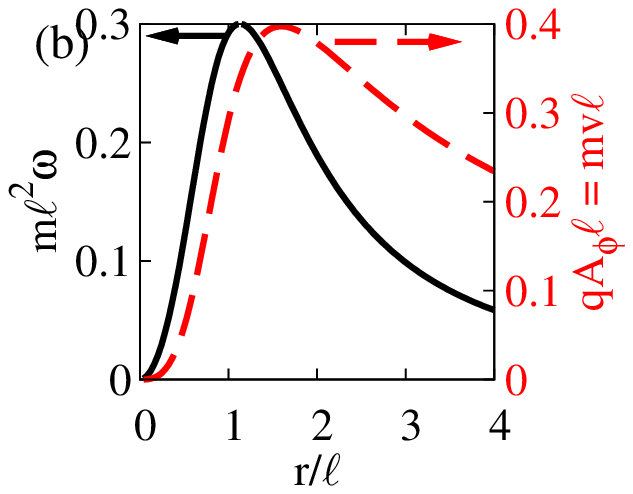}}
  \caption{Panel (a), microcavity sample ($\mu c$) placed between
    (imbalanced) anti-Helmholtz coils to induce a field
    $\vec{B}_{\text{real}}$. Panel (b), velocity and angular velocity vs
    radius for $\vec{A}_\text{synth}$ as discussed in the text.  }
  \label{fig:expt}
\end{figure}

In order to illustrate how this might work, one may consider the real
magnetic field produced by an imbalanced anti-Helmholtz configuration,
as illustrated in Fig.~\ref{fig:expt}(a), so that the magnetic field
in the microcavity has the form $\vec{B} = (\beta x, \beta y, B_z)$
for small in-plane coordinates $x,y$ with constant $B_z$.  Microcavity
polaritons typically involve heavy-hole excitons, so that polariton
polarizations $\pm 1$ imply electron and hole spins $(\mp 1/2, \pm
3/2)$.  As such, while a B field perpendicular to the microcavity
simply splits these polarization states, the in-plane field is more
complicated, as it mixes the polariton states with non-radiative
excitons with spins $(\pm 1/2, \pm 3/2)$.  Furthermore, depending on
the crystal symmetry and quantum well growth direction, the leading
order coupling between $\pm 3/2$ hole states may either be linear or
cubic in magnetic field~\cite{winkler:book}.  In order to illustrate
the basic idea, I will avoid these complications, and consider the
simplest situation, where a linear coupling exists\footnote{An
  alternative would be to use stress to produce an in-plane
  field~\cite{Balili2010}.}.  Adiabatically eliminating the
non-radiative excitons, the effective Hamiltonian for the polariton
polarization is $H = \lambda [ \ell^2 \sigma_z + r^2 (e^{2 i \phi}
\sigma_- + e^{-2i\phi} \sigma_+)]$ where $re^{i \phi} = x+iy$ and the
length $\ell$ encodes the ratio of in-plane and perpendicular fields.
One may then find the ground state $|\psi\rangle$, ensuring this is
smooth as $r\to 0$, and thus find $q \vec{A} = i \langle \psi|\nabla
\psi \rangle = (\hat{\phi}/r)(1-l^2/\sqrt{r^4+l^4})$.  This gauge
field is equivalent to a rotating frame with $ q
\vec{A}_{\text{synth}} = m \vec{\omega} \times \vec{r} $ hence
$\omega(r) = q A_{\phi} / m r$.  This is shown in
Fig.~\ref{fig:expt}(b).

Detection of the response to this rotating frame could potentially be
done by imaging the momentum and energy distribution of the
polaritons~\cite{deng10}.  If the condensate is concentrated around $r
\simeq \ell$, one has the maximum velocity, corresponding to an energy
shift for the normal component of $\Delta E_{\text{max}} = (1/2) m
v_{\text{max}}^2 = 0.08 / m \ell^2$.  For $\ell \simeq 0.5 \mu$m, and
$m=10^{-4}m_{\text{electron}}$ this corresponds to $0.2$meV.
Observing the differential shift of luminescence as one varies $\ell$
by varying $B_z$ could allow one to extract the superfluid fraction.

I now turn to discuss in more detail how the superfluid density can be
calculated in the Schwinger-Keldysh approach, and to explain the
origin of the vertex corrections in Fig.~\ref{fig:feynman} and the
form of the normal density in Eq.~(\ref{eq:4}).  Following the path
integral approach to the Schwinger-Keldysh formalism~\cite{altland_simons,*kamenev05}, the
response function can be written in terms of a generating functional as:
\begin{equation}
  \label{eq:5}
  \chi_{ij}(q) = 
  \frac{-i}{2} \frac{d^2 \ZZ[f,\theta]}{d f_i(q) d \theta_j(-q)}
  , \ \ 
  \ZZ =\!\!\int\!\mathcal{D}(\bar{\psi},\psi) e^{i (S+\delta S)}.
\end{equation}
Here, $S$ is the Keldysh action~\cite{altland_simons,*kamenev05} due
to the Hamiltonian in Eq.~(\ref{eq:3}) and the pump and decay terms $S
= \sum \bar{\psi}_{\vec{k}} \mathcal{D}_0^{-1} {\psi}_{\vec{k}^\prime}
- \iint dt d^2r \frac{U}{2}[ \bar{\psi}_q \bar{\psi}_{cl} (\psi_{cl}^2 + \psi_q^2) +
\text{H.c.} ]$ where the fields $\psi$ are written in the Keldysh
space for classical/quantum fields~\cite{altland_simons,*kamenev05},
and the retarded, advanced and Keldysh components of the inverse
Green's function $\mathcal{D}^{-1}_0$ are as discussed above.  The
source term $\delta S = \sum \bar{\psi}_{\vec{k}+\vec{q}} [\tau_1
f_i(q) + (\tau_3 + i\tau_2) \theta_i(q)] \gamma_i(2\vec{k}+\vec{q})
{\psi}_{\vec{k}}$, where Pauli matrices $\tau_i$ are in the Keldysh
space. The presence of two separate fields is necessary due to need to
calculate a normal ordered current (derivative with respect to
$\theta$) linearly dependent on a classical force (field $f$).

In the non-condensed state, one may immediately determine $\ZZ$ at
leading order by neglecting interactions, and performing the
(Gaussian) integration over fields $(\bar{\psi},\psi)_{cl,q}$. This
yields $ m \chi_{ij} = \rho_n \delta_{ij}$, with $\rho_n$ given by
Eq.~(\ref{eq:4}), where the Nambu structure in the normal state is
trivial.  When condensed, vertex corrections become important.  These
vertex corrections can be found by use of an ``honest saddle point''
of the partition function, i.e.\ determine the saddle point in the
presence of the source terms $f,\theta$, and then integrate out
fluctuations about this new saddle point.  One thus finds a form $\ZZ
\propto \exp\{ i S_0[f,\theta] - \frac{1}{2} \Tr \ln(1 + \mathcal{D}
A[f,\theta])\}$ where $S_0$ is the saddle point action, $\mathcal{D}$
are the Green's functions (for $f=\theta=0$) and $A$ is the self
energy due to the fields $f,\theta$.  Both $S_0$ and $A$ involve terms
arising from the shift of the saddle point field $\psi$ in the
presence of the source terms $f, \theta$, and thus both $S_0$ and $A$
have higher order contributions of $f,\theta$.  One may then expand
these terms to quadratic order in $f,\theta$ and evaluate $\chi_{ij}$
via Eq.~(\ref{eq:5}).

Taking derivatives with respect to $f, \theta$ (indicated by primes
and subscripts), one finds $\chi_{ij} = \frac{1}{2}
S^{\prime\prime}_{0,f_i,\theta_j} + \frac{i}{4} \Tr (\mathcal{D}
A^{\prime\prime}_{f_i \theta_j}) - \frac{i}{4} \Tr (\mathcal{D}
A^{\prime}_{f_i } \mathcal{D} A^{\prime}_{ \theta_j})$.  Comparing the
three terms in this expression to the diagrams in
Fig.~\ref{fig:feynman} the first diagram arises from the first term,
the second two diagrams arise from the second term, and the last three
diagrams from the third term.  After explicitly evaluating the shifts
to the saddle point, and the self energy $A$, one finds the explicit
form:
\begin{widetext}
    \begin{multline}
      \label{eq:6}
      m\chi_{ij}(q) = \frac{q_iq_j}{q^2} \left\{
        \frac{\Lambda}{U}
        + m\!\!\int\!\frac{d\epsilon_k}{2\pi} 
        \left[
          1 - \frac{i}{4} \int \frac{d\omega}{2\pi} \left(
            \text{Tr}\left[\left(2\sigma_0 +  \sigma_1\right)D^K_k\right]
            - \frac{2\Lambda}{\epsilon_q}
            \Tr(D^K_k \sigma_1) \right) \right]
      \right\}
      \\
      - m\!\!\int\! \frac{d\epsilon_k}{2\pi} \left[
        \frac{q_i q_j}{q^2} 2 \frac{\Lambda^2}{\epsilon_q} N_{22}
        -
        i \frac{q_i(q_j+2k_j)}{q^2} \Lambda N_{23}
        + 
        i \frac{(q_i+2k_i)q_j}{q^2} \Lambda N_{32}
        +
        \frac{(q_i+2k_i)(q_j+2k_j)}{4m} N_{33}
      \right]
    \end{multline}
\end{widetext}
where $N_{ab} = \int \frac{i d\omega}{8\pi} \Tr (D^R_{k+q} \sigma_a
D^K_k \sigma_b + D^K_{k+q} \sigma_a D^A_k \sigma_b)$ in terms of Pauli
matrices $\sigma_i$ in the Nambu space and $\Lambda = U|\psi_0|^2$.
The terms in Eq.~(\ref{eq:6}) are arranged in the same order as the
corresponding diagrams in Fig.~\ref{fig:feynman}.

A number of technical issues regarding regularization are worth noting.
Firstly, as is known elsewhere (see e.g.~\cite{wilson11} and refs.\
therein), there can be cases where it is necessary to return to the
discrete time coherent state path integral in order to correctly
incorporate causality in performing integrals. The current problem is
such a case, and the term arising from this is the $1$ on the first
line of Eq.~(\ref{eq:6}).  Secondly, in order to give an ultraviolet
finite expression, it is necessary to perform the standard T-matrix
regularization of the contact interaction: $U^{-1} \to U_{\text{eff}}^{-1} - m \int
\frac{d\epsilon_k}{2\pi} [2(\epsilon+\mu)]^{-1}$.  Thirdly, the
apparently singular terms involving $q_i q_j /q^2 \epsilon_q$ in fact cancel,
leaving only finite contributions as $q \to 0$.  One may verify that
Eq.~(\ref{eq:6}) recovers the expected equilibrium result in the
absence of pumping and decay.

One may note that the expression in Eq.~(\ref{eq:6}) does not
explicitly involve details of the pumping.  This is because
the only nonlinearity included being the interaction term $U$.  This
means that Eq.~(\ref{eq:6}) survives for general models of pumping and
decay, and so is more generic than the particular model of pumping used to
derive it.

In conclusion, the superfluid density of a non-equilibrium open
dissipative condensate need not vanish, despite the non-existence of a
Landau critical velocity.  This is because the poles of the response
function, which give the spectrum, 
do not uniquely determine the form of the response function
at zero frequency, which is the quantity that defines the superfluid
density.  Such a superfluid density could potentially be measured in a
polariton system by using real magnetic fields to engineer an
effective rotating frame.  The current-current response function can
be explicitly calculated using the ``honest saddle point'' approach.
Such an approach would also allow calculation of dynamical response
functions, allowing a more nuanced understanding of the distinctions
between static and dynamic superfluid phenomena in open dissipative
condensates.

\acknowledgments{I acknowledge helpful discussions with Austen
  Lamacraft and with Nigel Cooper, and funding from EPSRC grant
  EP/G004714/2}

\bibliography{non-eqbm-sf}

\begin{thebibliography}{28}%
\makeatletter
\providecommand \@ifxundefined [1]{%
 \@ifx{#1\undefined}
}%
\providecommand \@ifnum [1]{%
 \ifnum #1\expandafter \@firstoftwo
 \else \expandafter \@secondoftwo
 \fi
}%
\providecommand \@ifx [1]{%
 \ifx #1\expandafter \@firstoftwo
 \else \expandafter \@secondoftwo
 \fi
}%
\providecommand \natexlab [1]{#1}%
\providecommand \enquote  [1]{``#1''}%
\providecommand \bibnamefont  [1]{#1}%
\providecommand \bibfnamefont [1]{#1}%
\providecommand \citenamefont [1]{#1}%
\providecommand \href@noop [0]{\@secondoftwo}%
\providecommand \href [0]{\begingroup \@sanitize@url \@href}%
\providecommand \@href[1]{\@@startlink{#1}\@@href}%
\providecommand \@@href[1]{\endgroup#1\@@endlink}%
\providecommand \@sanitize@url [0]{\catcode `\\12\catcode `\$12\catcode
  `\&12\catcode `\#12\catcode `\^12\catcode `\_12\catcode `\%12\relax}%
\providecommand \@@startlink[1]{}%
\providecommand \@@endlink[0]{}%
\providecommand \url  [0]{\begingroup\@sanitize@url \@url }%
\providecommand \@url [1]{\endgroup\@href {#1}{\urlprefix }}%
\providecommand \urlprefix  [0]{URL }%
\providecommand \Eprint [0]{\href }%
\providecommand \doibase [0]{http://dx.doi.org/}%
\providecommand \selectlanguage [0]{\@gobble}%
\providecommand \bibinfo  [0]{\@secondoftwo}%
\providecommand \bibfield  [0]{\@secondoftwo}%
\providecommand \translation [1]{[#1]}%
\providecommand \BibitemOpen [0]{}%
\providecommand \bibitemStop [0]{}%
\providecommand \bibitemNoStop [0]{.\EOS\space}%
\providecommand \EOS [0]{\spacefactor3000\relax}%
\providecommand \BibitemShut  [1]{\csname bibitem#1\endcsname}%
\let\auto@bib@innerbib\@empty
%</preamble>
\bibitem [{\citenamefont {Leggett}(2006)}]{leggett06}%
  \BibitemOpen
  \bibfield  {author} {\bibinfo {author} {\bibfnamefont {A.~J.}\ \bibnamefont
  {Leggett}},\ }\href@noop {} {\emph {\bibinfo {title} {Quantum Liquids}}}\
  (\bibinfo  {publisher} {Oxford University Press},\ \bibinfo {year}
  {2006})\BibitemShut {NoStop}%
\bibitem [{\citenamefont {Deng}\ \emph {et~al.}(2010)\citenamefont {Deng},
  \citenamefont {Haug},\ and\ \citenamefont {Yamamoto}}]{deng10}%
  \BibitemOpen
  \bibfield  {author} {\bibinfo {author} {\bibfnamefont {H.}~\bibnamefont
  {Deng}}, \bibinfo {author} {\bibfnamefont {H.}~\bibnamefont {Haug}}, \ and\
  \bibinfo {author} {\bibfnamefont {Y.}~\bibnamefont {Yamamoto}},\ }\href
  {\doibase 10.1103/RevModPhys.82.1489} {\bibfield  {journal} {\bibinfo
  {journal} {Rev. Mod. Phys.}\ }\textbf {\bibinfo {volume} {82}},\ \bibinfo
  {pages} {1489} (\bibinfo {year} {2010})}\BibitemShut {NoStop}%
\bibitem [{\citenamefont {Klaers}\ \emph {et~al.}(2010)\citenamefont {Klaers},
  \citenamefont {Schmitt}, \citenamefont {Vewinger},\ and\ \citenamefont
  {Weitz}}]{Klaers10}%
  \BibitemOpen
  \bibfield  {author} {\bibinfo {author} {\bibfnamefont {J.}~\bibnamefont
  {Klaers}}, \bibinfo {author} {\bibfnamefont {J.}~\bibnamefont {Schmitt}},
  \bibinfo {author} {\bibfnamefont {F.}~\bibnamefont {Vewinger}}, \ and\
  \bibinfo {author} {\bibfnamefont {M.}~\bibnamefont {Weitz}},\ }\href
  {\doibase 10.1038/nature09567} {\bibfield  {journal} {\bibinfo  {journal}
  {Nature}\ }\textbf {\bibinfo {volume} {468}},\ \bibinfo {pages} {545}
  (\bibinfo {year} {2010})}\BibitemShut {NoStop}%
\bibitem [{\citenamefont {Falkenau}\ \emph {et~al.}()\citenamefont {Falkenau}
  \emph {et~al.}}]{Falkenau2011}%
  \BibitemOpen
  \bibfield  {author} {\bibinfo {author} {\bibfnamefont {M.}~\bibnamefont
  {Falkenau}} \emph {et~al.},\ }\href@noop {} {\ }\Eprint
  {http://arxiv.org/abs/arXiv:1102.0928} {arXiv:1102.0928} \BibitemShut
  {NoStop}%
\bibitem [{\citenamefont {Szyma\'nska}\ \emph {et~al.}(2006)\citenamefont
  {Szyma\'nska}, \citenamefont {Keeling},\ and\ \citenamefont
  {Littlewood}}]{szymanska06}%
  \BibitemOpen
  \bibfield  {author} {\bibinfo {author} {\bibfnamefont {M.~H.}\ \bibnamefont
  {Szyma\'nska}}, \bibinfo {author} {\bibfnamefont {J.}~\bibnamefont
  {Keeling}}, \ and\ \bibinfo {author} {\bibfnamefont {P.~B.}\ \bibnamefont
  {Littlewood}},\ }\href {\doibase 10.1103/PhysRevLett.96.230602} {\bibfield
  {journal} {\bibinfo  {journal} {Phys. Rev. Lett.}\ }\textbf {\bibinfo
  {volume} {96}},\ \bibinfo {pages} {230602} (\bibinfo {year}
  {2006})}\BibitemShut {NoStop}%
\bibitem [{\citenamefont {Wouters}\ and\ \citenamefont
  {Carusotto}(2006)}]{wouters06}%
  \BibitemOpen
  \bibfield  {author} {\bibinfo {author} {\bibfnamefont {M.}~\bibnamefont
  {Wouters}}\ and\ \bibinfo {author} {\bibfnamefont {I.}~\bibnamefont
  {Carusotto}},\ }\href {\doibase 10.1103/PhysRevB.74.245316} {\bibfield
  {journal} {\bibinfo  {journal} {Phys. Rev. B}\ }\textbf {\bibinfo {volume}
  {74}},\ \bibinfo {pages} {245316} (\bibinfo {year} {2006})}\BibitemShut
  {NoStop}%
\bibitem [{\citenamefont {Wouters}\ and\ \citenamefont
  {Carusotto}(2010)}]{wouters10:superfluid}%
  \BibitemOpen
  \bibfield  {author} {\bibinfo {author} {\bibfnamefont {M.}~\bibnamefont
  {Wouters}}\ and\ \bibinfo {author} {\bibfnamefont {I.}~\bibnamefont
  {Carusotto}},\ }\href {\doibase 10.1103/PhysRevLett.105.020602} {\bibfield
  {journal} {\bibinfo  {journal} {Phys. Rev. Lett.}\ }\textbf {\bibinfo
  {volume} {105}},\ \bibinfo {pages} {020602} (\bibinfo {year}
  {2010})}\BibitemShut {NoStop}%
\bibitem [{\citenamefont {Wouters}\ and\ \citenamefont
  {Savona}(2010)}]{wouters10:vortices}%
  \BibitemOpen
  \bibfield  {author} {\bibinfo {author} {\bibfnamefont {M.}~\bibnamefont
  {Wouters}}\ and\ \bibinfo {author} {\bibfnamefont {V.}~\bibnamefont
  {Savona}},\ }\href {\doibase 10.1103/PhysRevB.81.054508} {\bibfield
  {journal} {\bibinfo  {journal} {Phys. Rev. B}\ }\textbf {\bibinfo {volume}
  {81}},\ \bibinfo {pages} {054508} (\bibinfo {year} {2010})}\BibitemShut
  {NoStop}%
\bibitem [{\citenamefont {Lagoudakis}\ \emph {et~al.}(2008)\citenamefont
  {Lagoudakis} \emph {et~al.}}]{Lagoudakis2008}%
  \BibitemOpen
  \bibfield  {author} {\bibinfo {author} {\bibfnamefont {K.~G.}\ \bibnamefont
  {Lagoudakis}} \emph {et~al.},\ }\href {\doibase 10.1038/nphys1051} {\bibfield
   {journal} {\bibinfo  {journal} {Nature Phys.}\ }\textbf {\bibinfo {volume}
  {4}},\ \bibinfo {pages} {706} (\bibinfo {year} {2008})}\BibitemShut {NoStop}%
\bibitem [{\citenamefont {Amo}\ \emph {et~al.}(2009{\natexlab{a}})\citenamefont
  {Amo} \emph {et~al.}}]{Amo2009}%
  \BibitemOpen
  \bibfield  {author} {\bibinfo {author} {\bibfnamefont {A.}~\bibnamefont
  {Amo}} \emph {et~al.},\ }\href {\doibase 10.1038/nature07640} {\bibfield
  {journal} {\bibinfo  {journal} {Nature}\ }\textbf {\bibinfo {volume} {457}},\
  \bibinfo {pages} {291} (\bibinfo {year} {2009}{\natexlab{a}})}\BibitemShut
  {NoStop}%
\bibitem [{\citenamefont {Amo}\ \emph {et~al.}(2009{\natexlab{b}})\citenamefont
  {Amo} \emph {et~al.}}]{Amo2009a}%
  \BibitemOpen
  \bibfield  {author} {\bibinfo {author} {\bibfnamefont {A.}~\bibnamefont
  {Amo}} \emph {et~al.},\ }\href {\doibase 10.1038/nphys1364} {\bibfield
  {journal} {\bibinfo  {journal} {Nature Phys.}\ }\textbf {\bibinfo {volume}
  {5}},\ \bibinfo {pages} {805} (\bibinfo {year}
  {2009}{\natexlab{b}})}\BibitemShut {NoStop}%
\bibitem [{\citenamefont {Sanvitto}\ \emph {et~al.}(2010)\citenamefont
  {Sanvitto} \emph {et~al.}}]{Krizhanovskii2010}%
  \BibitemOpen
  \bibfield  {author} {\bibinfo {author} {\bibfnamefont {D.}~\bibnamefont
  {Sanvitto}} \emph {et~al.},\ }\href {\doibase 10.1038/nphys1668} {\bibfield
  {journal} {\bibinfo  {journal} {Nature Phys.}\ }\textbf {\bibinfo {volume}
  {6}},\ \bibinfo {pages} {527} (\bibinfo {year} {2010})}\BibitemShut {NoStop}%
\bibitem [{\citenamefont {Altland}\ and\ \citenamefont
  {Simons}(2010)}]{altland_simons}%
  \BibitemOpen
  \bibfield  {author} {\bibinfo {author} {\bibfnamefont {A.}~\bibnamefont
  {Altland}}\ and\ \bibinfo {author} {\bibfnamefont {B.~D.}\ \bibnamefont
  {Simons}},\ }\href@noop {} {\emph {\bibinfo {title} {Condensed Matter Field
  Theory}}},\ \bibinfo {edition} {2nd}\ ed.\ (\bibinfo  {publisher} {Cambridge
  University Press},\ \bibinfo {year} {2010})\BibitemShut {NoStop}%
\bibitem [{\citenamefont {Kamenev}(2005)}]{kamenev05}%
  \BibitemOpen
  \bibfield  {author} {\bibinfo {author} {\bibfnamefont {A.}~\bibnamefont
  {Kamenev}},\ }in\ \href@noop {} {\emph {\bibinfo {booktitle} {Nanophysics:
  Coherence and transport}}},\ \bibinfo {series} {Les Houches}, Vol.\ \bibinfo
  {volume} {LXXXI},\ \bibinfo {editor} {edited by\ \bibinfo {editor}
  {\bibfnamefont {H.}~\bibnamefont {Bouchiat}} \emph {et~al.}}\ (\bibinfo
  {publisher} {Elsevier},\ \bibinfo {address} {Amsterdam},\ \bibinfo {year}
  {2005})\ p.\ \bibinfo {pages} {177}\BibitemShut {NoStop}%
\bibitem [{Note1()}]{Note1}%
  \BibitemOpen
  \bibinfo {note} {To avoid ultraviolet divergence, a regularization $[\kappa -
  i p(\omega )] \to [\kappa - i p(\omega )] i \Gamma /(\omega +i\Gamma )$ is
  also required. $\Gamma $ is assumed large compared to all other energy
  scales.}\BibitemShut {Stop}%
\bibitem [{\citenamefont {Wouters}\ and\ \citenamefont
  {Savona}()}]{wouters10:relax}%
  \BibitemOpen
  \bibfield  {author} {\bibinfo {author} {\bibfnamefont {M.}~\bibnamefont
  {Wouters}}\ and\ \bibinfo {author} {\bibfnamefont {V.}~\bibnamefont
  {Savona}},\ }\href@noop {} {}\Eprint {http://arxiv.org/abs/arXiv:1007.5453}
  {arXiv:1007.5453} \BibitemShut {NoStop}%
\bibitem [{\citenamefont {Griffin}(1994)}]{griffin94}%
  \BibitemOpen
  \bibfield  {author} {\bibinfo {author} {\bibfnamefont {A.}~\bibnamefont
  {Griffin}},\ }\href@noop {} {\emph {\bibinfo {title} {Excitations in a
  {B}ose-Condensed Liquid}}}\ (\bibinfo  {publisher} {Cambridge University
  Press, Cambridge},\ \bibinfo {year} {1994})\BibitemShut {NoStop}%
\bibitem [{\citenamefont {Shelykh}\ \emph {et~al.}(2010)\citenamefont {Shelykh}
  \emph {et~al.}}]{shelykh10:review}%
  \BibitemOpen
  \bibfield  {author} {\bibinfo {author} {\bibfnamefont {I.~A.}\ \bibnamefont
  {Shelykh}} \emph {et~al.},\ }\href
  {http://stacks.iop.org/0268-1242/25/i=1/a=013001} {\bibfield  {journal}
  {\bibinfo  {journal} {Semicond. Sci. Technol.}\ }\textbf {\bibinfo {volume}
  {25}},\ \bibinfo {pages} {013001} (\bibinfo {year} {2010})}\BibitemShut
  {NoStop}%
\bibitem [{\citenamefont {Cooper}\ and\ \citenamefont
  {Hadzibabic}(2010)}]{Cooper2010}%
  \BibitemOpen
  \bibfield  {author} {\bibinfo {author} {\bibfnamefont {N.~R.}\ \bibnamefont
  {Cooper}}\ and\ \bibinfo {author} {\bibfnamefont {Z.}~\bibnamefont
  {Hadzibabic}},\ }\href {\doibase 10.1103/PhysRevLett.104.030401} {\bibfield
  {journal} {\bibinfo  {journal} {Phys. Rev. Lett.}\ }\textbf {\bibinfo
  {volume} {104}},\ \bibinfo {pages} {030401} (\bibinfo {year}
  {2010})}\BibitemShut {NoStop}%
\bibitem [{\citenamefont {Dalibard}\ \emph {et~al.}()\citenamefont {Dalibard},
  \citenamefont {Gerbier}, \citenamefont {Juzeli{\~{u}}nas},\ and\
  \citenamefont {\"Ohberg}}]{dalibard}%
  \BibitemOpen
  \bibfield  {author} {\bibinfo {author} {\bibfnamefont {J.}~\bibnamefont
  {Dalibard}}, \bibinfo {author} {\bibfnamefont {F.}~\bibnamefont {Gerbier}},
  \bibinfo {author} {\bibfnamefont {G.}~\bibnamefont {Juzeli{\~{u}}nas}}, \
  and\ \bibinfo {author} {\bibfnamefont {P.}~\bibnamefont {\"Ohberg}},\
  }\href@noop {} {}\Eprint {http://arxiv.org/abs/arXiv:1008.5378}
  {arXiv:1008.5378} \BibitemShut {NoStop}%
\bibitem [{\citenamefont {Cooper}(2011)}]{Cooper2011}%
  \BibitemOpen
  \bibfield  {author} {\bibinfo {author} {\bibfnamefont {N.~R.}\ \bibnamefont
  {Cooper}},\ }\href {\doibase 10.1103/PhysRevLett.106.175301} {\bibfield
  {journal} {\bibinfo  {journal} {Phys. Rev. Lett.}\ }\textbf {\bibinfo
  {volume} {106}},\ \bibinfo {pages} {175301} (\bibinfo {year}
  {2011})}\BibitemShut {NoStop}%
\bibitem [{\citenamefont {Hafezi}\ \emph {et~al.}(2007)\citenamefont {Hafezi},
  \citenamefont {S\"orensen}, \citenamefont {Demler},\ and\ \citenamefont
  {Lukin}}]{Hafezi2007}%
  \BibitemOpen
  \bibfield  {author} {\bibinfo {author} {\bibfnamefont {M.}~\bibnamefont
  {Hafezi}}, \bibinfo {author} {\bibfnamefont {A.}~\bibnamefont {S\"orensen}},
  \bibinfo {author} {\bibfnamefont {E.}~\bibnamefont {Demler}}, \ and\ \bibinfo
  {author} {\bibfnamefont {M.}~\bibnamefont {Lukin}},\ }\href {\doibase
  10.1103/PhysRevA.76.023613} {\bibfield  {journal} {\bibinfo  {journal} {Phys.
  Rev. A}\ }\textbf {\bibinfo {volume} {76}},\ \bibinfo {pages} {023613}
  (\bibinfo {year} {2007})}\BibitemShut {NoStop}%
\bibitem [{\citenamefont {Koch}\ \emph {et~al.}(2010)\citenamefont {Koch},
  \citenamefont {Houck}, \citenamefont {{Le Hur}},\ and\ \citenamefont
  {Girvin}}]{Koch2010}%
  \BibitemOpen
  \bibfield  {author} {\bibinfo {author} {\bibfnamefont {J.}~\bibnamefont
  {Koch}}, \bibinfo {author} {\bibfnamefont {A.}~\bibnamefont {Houck}},
  \bibinfo {author} {\bibfnamefont {K.}~\bibnamefont {{Le Hur}}}, \ and\
  \bibinfo {author} {\bibfnamefont {S.}~\bibnamefont {Girvin}},\ }\href
  {\doibase 10.1103/PhysRevA.82.043811} {\bibfield  {journal} {\bibinfo
  {journal} {Phys. Rev. A}\ }\textbf {\bibinfo {volume} {82}},\ \bibinfo
  {pages} {043811} (\bibinfo {year} {2010})}\BibitemShut {NoStop}%
\bibitem [{\citenamefont {Umucalilar}\ and\ \citenamefont
  {Carusotto}()}]{Umucalilar2011}%
  \BibitemOpen
  \bibfield  {author} {\bibinfo {author} {\bibfnamefont {R.~O.}\ \bibnamefont
  {Umucalilar}}\ and\ \bibinfo {author} {\bibfnamefont {I.}~\bibnamefont
  {Carusotto}},\ }\href@noop {} {\ }\Eprint
  {http://arxiv.org/abs/arXiv:1104.4071} {arXiv:1104.4071} \BibitemShut
  {NoStop}%
\bibitem [{\citenamefont {Winkler}(2003)}]{winkler:book}%
  \BibitemOpen
  \bibfield  {author} {\bibinfo {author} {\bibfnamefont {R.}~\bibnamefont
  {Winkler}},\ }\href@noop {} {\emph {\bibinfo {title} {Spin-Orbit Coupling
  Effects in Two-Dimensional Electron and Hole Systems}}},\ \bibinfo {series}
  {Springer Tracts in Modern Physics}\ No.\ \bibinfo {number} {191}\ (\bibinfo
  {publisher} {Springer},\ \bibinfo {address} {Berlin},\ \bibinfo {year}
  {2003})\BibitemShut {NoStop}%
\bibitem [{Note2()}]{Note2}%
  \BibitemOpen
  \bibinfo {note} {An alternative would be to use stress to produce an in-plane
  field~\cite {Balili2010}.}\BibitemShut {Stop}%
\bibitem [{\citenamefont {Wilson}\ and\ \citenamefont
  {Galitski}(2011)}]{wilson11}%
  \BibitemOpen
  \bibfield  {author} {\bibinfo {author} {\bibfnamefont {J.~H.}\ \bibnamefont
  {Wilson}}\ and\ \bibinfo {author} {\bibfnamefont {V.}~\bibnamefont
  {Galitski}},\ }\href {\doibase 10.1103/PhysRevLett.106.110401} {\bibfield
  {journal} {\bibinfo  {journal} {Phys. Rev. Lett.}\ }\textbf {\bibinfo
  {volume} {106}},\ \bibinfo {pages} {110401} (\bibinfo {year}
  {2011})}\BibitemShut {NoStop}%
\bibitem [{\citenamefont {Balili}\ \emph {et~al.}(2010)\citenamefont {Balili},
  \citenamefont {Nelsen}, \citenamefont {Snoke}, \citenamefont {Pfeiffer},\
  and\ \citenamefont {West}}]{Balili2010}%
  \BibitemOpen
  \bibfield  {author} {\bibinfo {author} {\bibfnamefont {R.}~\bibnamefont
  {Balili}}, \bibinfo {author} {\bibfnamefont {B.}~\bibnamefont {Nelsen}},
  \bibinfo {author} {\bibfnamefont {D.~W.}\ \bibnamefont {Snoke}}, \bibinfo
  {author} {\bibfnamefont {L.}~\bibnamefont {Pfeiffer}}, \ and\ \bibinfo
  {author} {\bibfnamefont {K.}~\bibnamefont {West}},\ }\href {\doibase
  10.1103/PhysRevB.81.125311} {\bibfield  {journal} {\bibinfo  {journal} {Phys.
  Rev. B}\ }\textbf {\bibinfo {volume} {81}},\ \bibinfo {pages} {125311}
  (\bibinfo {year} {2010})}\BibitemShut {NoStop}%
\end{thebibliography}%

\end{document}